\documentclass[a4paper]{PoS}

\usepackage{amsmath}
\usepackage{amssymb}
\usepackage{amsthm}

\usepackage{graphicx}
\usepackage{ctable}
\def\apj{ApJ, }%
\def\aap{A\&A, }%
\def\apjl{ApJ, }%
\def\mnras{MNRAS, }%
\def\pasj{PASJ, }

\def\psrb{PSR\,B1259$-$63/LS 2883}

\def\gammaray{$\gamma$-ray}
\def\gammarays{$\gamma$-rays}
\def\gaga{$\gamma\gamma$}
\def\hess{H.E.S.S.}

\def\fermi{\textit{Fermi}-LAT}

\def\deg{$^{\circ}$}

\title{Role of the disk environment in the gamma-ray emission from the binary system PSR B1259-63/LS 2883}

\ShortTitle{Disk environment in PSR B1259-63/LS 2883}

\author{\speaker{Iurii Sushch}\\%\thanks{A footnote may follow.}\\
        Centre for Space Research, North-West University, Potchefstroom 2520, South Africa\\
        Astronomical Observatory of Ivan Franko National University of L'viv, vul. Kyryla i Methodia, 8, L'viv 79005, Ukraine\\
        E-mail: \email{iurii.sushch@nwu.ac.za}}

\author{Markus B\"ottcher\footnote{The work of M.B. is supported by the South African Research Chair Initiative (SARChI) of the South African National Research Foundation and the Department of Science and Technology. Disclaimer: Any opinion, finding and conclusion or recommendation expressed in this material is that of the authors and the NRF does not accept any liability in this regard.}\\
Centre for Space Research, North-West University, Potchefstroom 2520, South Africa}

\abstract{PSR B1259-63/LS 2883 is a very high energy (VHE; E > 100 GeV) gamma-ray emitting binary consisting of a 48 ms pulsar orbitting around a Be star with a period of 3.4 years. The Be star features a circumstellar disk which is inclined with respect to the orbit in such a way that the pulsar crosses it twice every orbit. The circumstellar disk provides an additional field of target photons which may contribute to inverse Compton scattering and gamma-gamma absorption, leaving a characteristic imprint in the observed spectrum and light curve of the high energy emission. We study the signatures of Compton-supported, VHE gamma-ray induced pair cascades in the circumstellar disc of the Be star and their possible contribution to the GeV flux. We also study a possible impact of the gamma-gamma absorption in the disk on the observed TeV light curve. We show that the cumulative absorption of VHE gamma-rays in stellar and disk photon fields can explain the modulation of the flux at the periastron passage.}

\FullConference{The 34th International Cosmic Ray Conference,\\
		30 July- 6 August, 2015\\
		The Hague, The Netherlands}

\begin{document}

\section{Introduction}
\label{intro}
\psrb\ is a member of the small class of very high energy (VHE; $E>100$ GeV) \gammaray\ binaries which comprises only 
five known objects. \psrb\ is unique for being the only \gammaray\ binary for which the 
compact object is clearly identified as a pulsar. This pulsar with a spin period of $\simeq 48$~ms and a spin-down 
luminosity of $\simeq 8 \times 10^{35}$~erg~s$^{-1}$ is moving in a highly eccentric ($e = 0.87$) 
orbit around a massive Be star with a period of $P_{\mathrm{orb}} = 3.4$ years (1237 days) 
\cite{1992MNRAS.255..401J, 1992ApJ...387L..37J}. The companion star LS 2883 has a luminosity 
of $L_{\ast} = 2.3 \times 10^{38}$~erg~s$^{-1}$. Because of the fast rotation of the star, it has an oblate shape 
with an equatorial radius of $R_{\mathrm{eq}} = 9.7 R_{\odot}$ and a polar radius of $R_{\mathrm{pole}} = 8.1 R_{\odot}$, 
which in turn leads to a strong gradient of the surface temperature from $T_{\mathrm{eq}} \simeq 27,500$\,K at the equator 
to $T_{\mathrm{pole}} \simeq 34,000$\,K at the poles \cite{2011ApJ...732L..11N}. 

The Be star features an equatorial 
circumstellar disk, which is believed to be inclined with respect to the pulsar's orbit (see e.g. \cite{2011ApJ...732L..11N}), so that the pulsar crosses the 
disk twice each orbit. The circumstellar disk of a Be star is a decretion disk with an enhanced stellar outflow formed around the star. 
As shown in optical interferometry observations, these disks are symmetrical with respect to the star's rotation 
axis (see e.g. \cite{1994A&A...283L..13Q}). 
Circumstellar disks generate excess infrared (IR) emission produced through free-free and 
free-bound radiation, providing an additional IR photon field to the blackbody flux from the optical star.

The dense medium of the disk is believed to play an essential role in the resulting emission from the system. The 
position of the disk can be localized based on the disappearance of the pulsed radio emission from the pulsar. 
The observed radio emission far from periastron consists only of the pulsed component \cite{1999MNRAS.302..277J, 2005MNRAS.358.1069J}, 
but closer to periastron, at about $t_{\mathrm{p}} - 100$\,d ($t_{\mathrm{p}}$ is the time of periastron), the intensity of the 
pulsed emission starts to decrease and completely disappears at about $t_{\mathrm{p}} - 20$\,d. It then re-appears at around $t_{\mathrm{p}} + 15$\,d. 
This eclipse of the pulsed emission is believed to be caused by the circumstellar disk.
It is accompanied by an increase of the transient unpulsed radio flux beginning at 
$\sim t_{\mathrm{p}}-30$\,d and reaching its maximum at $\sim t_{\mathrm{p}}-10$\,d. This is followed by a decrease around the 
periastron passage and a second peak at about $t_{\mathrm{p}} + 20$\,d \cite{1999MNRAS.302..277J, 2005MNRAS.358.1069J, 2014MNRAS.439..432C}.
A similar behavior is observed also for the unpulsed X-ray emission. Close to the periastron passage the X-ray emission features 
two peaks at around 20 days before and after periastron with flux levels 10 -- 20 times 
higher than during apastron, and a decrease of the emission at the time of the periastron passage itself. 
The X-ray data is very similar from orbit to orbit repeating the shape of the light curve 
very well (\cite{2014MNRAS.439..432C} and references therein). These peaks might be connected 
to the crossing of the disk environment.

At TeV energies the source was observed by \hess\ around four periastron passages in 2004 \cite{2005A&A...442....1A}, 2007 
\cite{2009A&A...507..389A}, 2010 \cite{2013A&A...551A..94H}, and 2104 (results not available yet). The TeV emission from the source shows a variable behavior 
around the periastron passage. Although the exposure of the source is not sufficient to draw firm conclusions, the combined 
light curves from all three observing campaigns show a hint of two asymmetrical peaks before and after periastron, which 
roughly coincide in time with the peaks of the emission in the radio and X-ray bands, and a decrease of the flux at periastron. 
The shape of the light curve might be a result of the interaction of the pulsar with the circumstellar disk.

First observations at GeV energies conducted by \fermi\ took place around the 2010 periastron passage revealing 
quite unexpected results \cite{2011ApJ...736L..11A, 2011ApJ...736L..10T}. \fermi\ detected a low flux from the source 
close to periastron with a subsequent dissapearance of the 
source after periastron followed by a sudden flare (with $\sim10$ times the pre-periastron flux level) 30 days after periastron. 
This flare lasted for about 7 weeks without any obvious counterparts at other energy bands. The flare is shifted in time 
with respect to the post-periastron peak at other energy bands. The nature of the flare is still not understood, but 
several possible explanations have been discussed in the literature \cite{2012ApJ...752L..17K, 2008MNRAS.387...63B, 2010A&A...516A..18D, 2012ApJ...753..127K, 2012ApJ...753..127K, 2013A&A...557A.127D}. New observations around the 2014 periastron passage revealed the GeV flare at the similar 
orbital phase establishing the repetitive behaviour of this phenomenon \cite{2015ApJ...798L..26T}. However, new observations 
did not show any significant emission close to periastron \cite{2015ApJ...798L..26T}.

The circumstellar disk of the companion star plays a crucial role in the variability of the \gammaray\ emission from 
the system. The dense disk photon field should significantly contribute to the target photons for inverse Compton scattering from the source, enhancing 
the observed TeV emission. The abrupt change of density at the pulsar's entrance and escape from the disk should also cause 
a change of the shape of the shock between the pulsar wind and stellar environment, which might be the reason of such spectacular 
events as the GeV flare. The high density of seed photons should also increase the opacity for \gaga-absorption, followed by electron-positron 
pair production, which in turn might scatter again the disk photon field, generating secondary \gammarays. This cascade 
process may cause the re-emission of the TeV IC flux at lower GeV energies. Moreover, because of the deflection of electrons 
and positrons in the magnetic field, secondary \gammarays\ can be re-emitted into randomized directions. Therefore, even if 
the primary VHE photon was emitted into the direction opposite to the line of sight to the observer, it can still contribute to the observable 
flux at lower GeV energies through the cascade emission. The pair cascade emission in binary systems caused by the interaction 
of the primary very high energy photons with the stellar photons was studied in detail for several cases 
(see e.g. \cite{2005MNRAS.356..711S, 2010A&A...519A..81C} and references therein). These studies showed that, although environments of the binary systems 
fulfill all the requirements for effective pair cascading, the resulting spectrum is in conflict with the GeV data observed by \fermi, 
since the expected cascade emission peaks in the \fermi\ energy band with very constraining upper limits. The \gaga-absorption of the 
TeV \gammarays\ by stellar photons was also suggested as a possible explanation of the variability of the TeV flux across the orbit 
\cite{2006A&A...451....9D}. The idea is that at periastron, when the pulsar is at the shortest distance to the star, the absorption 
should be the most effective, which would provide the decrease of the flux. However, \cite{2006A&A...451....9D} showed that absorption only 
cannot explain the TeV light curve.

\section{Cascade Emission in the Circumstellar Disk}

We used the cascade Monte Carlo code of \cite{2010ApJ...717..468R, 2012ApJ...750...26R} to calculate the angle-dependent cascade spectra for a variety 
of different input parameter sets within a parameter space motivated by the known properties of the Be star and its 
circumstellar disk in the \psrb\ binary system. In these calculations we assume that the pulsar is a point-like source emitting 
isotropically in all directions. This assumption is reasonable if the wind termination shock at which leptons injected by the pulsar 
are efficiently accelerated and isotropized is close to the pulsar and the density of target seed photons is high enough to lead 
to IC scattering in the direct vicinity of the shock, which is the case for \psrb\ \cite{2014JHEAp...3...18S}. We approximate the spectrum of the 
VHE \gammarays\ as generated by IC scattering by electrons as a power law distribution with an exponential cut-off. In our simulations we consider a 
mono-directional beam of photons to isolate all the geometrical effects. The source would emit the same beam of photons in every 
direction, and in the case of efficient cascading even those photons emitted in the opposite direction from the observer 
can contribute to the resulting observable spectrum. The orientation of the magnetic field would be different for primary photons 
emitted in different directions. The disk is approximated by a grey-body with the energy density $u_{\mathrm{d}}$ and temperature 
$T_{\mathrm{d}} = 0.6 \, T_{\ast}$ \cite{2011PASJ...63..893O, 2012ApJ...750...70T, 2012MNRAS.426.3135V} and the shape of the disk is 
approximated by a cuboid whose side lengths are chosen in a way to correspond to the real size of the disk. All the details on model assumptions 
and their justification can be found in \cite{2014JHEAp...3...18S}.

\begin{figure}
\centering
\resizebox{\hsize}{!}{\includegraphics{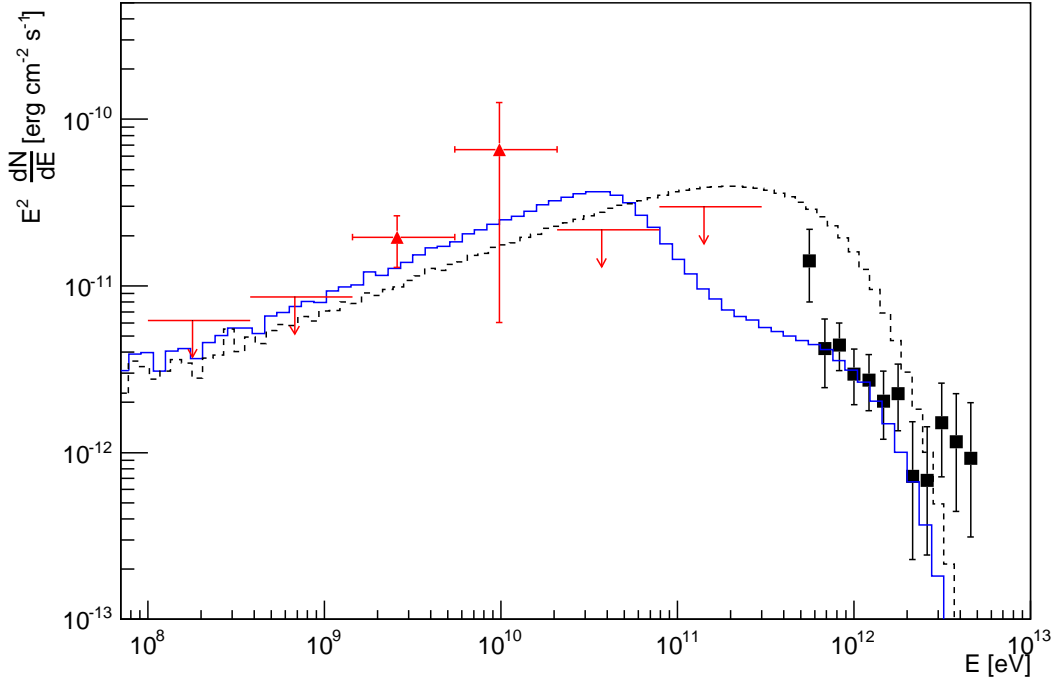}}
\caption{Comparison with the data. The black dashed line represents the intrinsic \gammaray\ spectrum. 
The blue solid line represents the spectrum of the forward emission ($0.98\leq\mu\leq1$) 
modified by gamma-gamma absorption and cascading. Assumed parameters: $B = 0.1$\,G, $\beta = 10$, $u_{\mathrm{d}} = 20$ erg/cm$^{3}$. 
Black squares represent the 2011 H.E.S.S. data \cite{2013A&A...551A..94H} and red triangles 
show the GeV data close to the 2010 periastron passage as reported by \cite{2011ApJ...736L..10T}.}
\label{cas_data}
\end{figure}

The dependence of the cascade emission on the viewing angle, magnetic field strength and orientation, energy density in the disk, and source location 
within the disk were discussed in detail in \cite{2014JHEAp...3...18S}. It was also shown that the GeV flare is very unlikely to be genereated by the 
cascade emission. However, cascade emission might still contribute to the faint GeV emission observed close to periastron during the periastron passage in 2010. 
The spectrum presented by \cite{2011ApJ...736L..10T} is consistent with a substantial contribution from cascade 
emission. The energy of the peak in the spectrum of the cascade \gammarays\ emitted in the forward direction 
only slightly depends on the energy density and magnetic field \cite{2014JHEAp...3...18S} and is in the energy 
range of $15-30$ GeV. Figure\,\ref{cas_data} shows the comparison of the GeV-TeV data with the simulation results of the 
cascade Compton radiation from the primary photon beam directed towards the observer emitted in the forward direction 
into a cone with an opening angle 11\deg\ ($0.98\leq\mu\leq1$). This corresponds to the cascade emission radiated in 
the direction to the observer from the primary \gammarays\ emitted within the cone of the same opening angle assuming 
that the magnetic field affects the cascades initiated by every primary photon identically. This assumption is valid 
for sufficiently small values of the opening angle, such that the orientation 
of the magnetic field with respect to the direction of the primary photon does not change much and its influence on 
the shape of the cascade emission spectrum is negligible. The simulation was performed for $u_{\mathrm{d}} = 20$ erg/cm$^{3}$, 
magnetic field $B = 0.1$ G with $\beta = B_{\mathrm{x}}/B_{\mathrm{y}} = 10$ (\textbf{B} lies in the \emph{xy-}plane). 
We decided to chose a magnetic-field direction closely aligned with the primary photon beam (and thus, the direction to the observer). 
In this way, the magnetic field has very little influence on the cascade development and on the resulting Compton emission. 
Otherwise, the cascade radiation emitted in other directions 
can become higher than the forward emission and a cone with a bigger opening angle should be considered, which in turn requires an accurate 
calculation of the magnetic field orientation for every primary photon. 

However, the spectrum of the GeV emission during the "low-flux" period 
derived by an independent group features a completely different spectral shape \cite{2011ApJ...736L..11A}. It reveals a significant flux at 
$0.1 - 1$\,GeV with no emission above $1$\,GeV. In this case the spectrum rejects the possibility of a significant cascade contribution, 
because even a very small level of absorption of the TeV \gammarays\ and re-emission at lower energies would violate the \fermi\ upper limit 
in $1-100$\,GeV energy band. The \fermi\ upper limit in this case constrains the energy density in the disk to 
$u_{\mathrm{d}}\lesssim 8$ erg/cm$^{3}$ \cite{2014JHEAp...3...18S}. Unfortunately, no significant emission close to periastron was detected 
during the 2014 periastron passage \cite{2015ApJ...798L..26T}, thus, the spectral discrepancy remains unresolved. 

It should be noted though that the simulations presented above are performed considering only the disk radiation field and 
neglecting the stellar radiation field. Therefore, the comparison of the model with 
the "periastron" emission presented on Fig. \ref{cas_data} should be treated only as a hint that the GeV emission may be 
explained by the cascade radiation (even if we consider only disk photons). This implies that taking into account stellar 
photons might probably provide a better fit of the data and decrease the estimate (upper limit) of the disk energy density. A detailed study 
of the cascade development in the system which would properly account for all relevant effects is in progress.

\section{Impact of the \gaga-Absorption on the Observed TeV Light Curve}

\begin{figure}
\centering
\resizebox{\hsize}{!}{\includegraphics{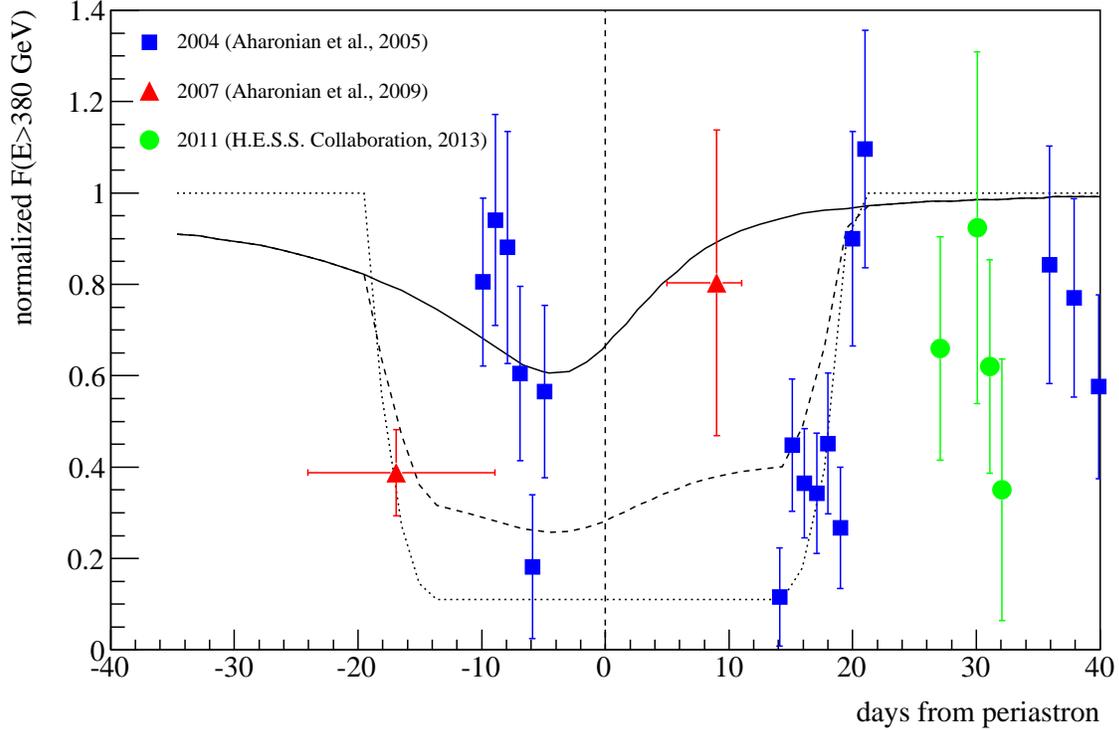}}
\caption{VHE $\gamma$-ray flux modulation caused by \gaga-absorption in \psrb. The solid line represents the normalized integrated flux above 380 GeV 
after absorption caused by the stellar radiation field as calculated in \cite{2006A&A...451....9D}. The dashed line shows the same flux 
taking into account also the absorption caused by the disk radiation field 
assuming the \gaga\ optical depth is constant above 380\,GeV. The dotted line 
represents the normalized integrated flux above 380\,GeV after absorption 
caused by the circumstellar disk only taking into account the dependancy 
of the \gaga\ optical depth on energy. See the main text for details on the 
geometry and other parameters.
The data points represent the combined H.E.S.S. 
light curve from three observed periastron passages as reported in \cite{2013A&A...551A..94H}. The integrated flux above 1\,TeV is extrapolated 
down to 380\,GeV using a photon index of $2.8$. The same flux normalisation factor as in \cite{2006A&A...451....9D} of $10^{-11}$\,cm$^{-2}$\,s$^{-1}$ 
is used.}
\label{tev_lc}
\end{figure}

The absorption of the \gammaray\ emission caused by the interaction with the disk radiation field might have 
a significant impact at the observed TeV light curve. The resulting integrated flux above 380\,GeV after 
absorption caused by the stellar radiation field was calculated in \cite{2006A&A...451....9D}. It was shown that the absorption due to the 
stellar radiation plays only a minor role in the observed variability. However, the disk radiation was not considered in those 
calculations. Using the estimate of the resulting integrated flux after the absorption on stellar photons only, $F_{\mathrm{s}}$, 
calculated in \cite{2006A&A...451....9D} we calculated the resulting flux after the total absorption due to both stellar and disk 
radiation fields as
\begin{equation}
F(E>380\,\mathrm{GeV}) \simeq F_{\mathrm{s}}(E>380\,\mathrm{GeV}) e^{-\bar{\tau}_{\gamma\gamma}} = F_{\mathrm{s}}(E>380\,\mathrm{GeV})\frac{\int_{\epsilon_{\gamma,\,\mathrm{min}}}^{\epsilon_{\gamma,\,\mathrm{max}}}e^{-\tau_{\gamma\gamma,\,\mathrm{d}}(\epsilon_\gamma)} d\epsilon_{\gamma}} {\epsilon_{\gamma,\,\mathrm{max}} - \epsilon_{\gamma,\,\mathrm{min}}},
\end{equation}
assuming the optical depth of the disk radiation field  $\tau_{\gamma\gamma,\,\mathrm{d}}(\epsilon)$ only weekly depends on energy ($\epsilon_\gamma = E_\gamma/(m_{\mathrm{e}}c^2)$ is the normalised \gammaray\ energy) in the energy range $E_{\gamma,\,\mathrm{min}} = 380$\,GeV and $E_{\gamma,\,\mathrm{max}} = 10$\,TeV and can be replaced by a constant averaged value $\bar{\tau}_{\gamma\gamma}$. For these calculations we used the same orbital parameters as used 
in \cite{2006A&A...451....9D}: orbital period $P_{\mathrm{orb}} = 1236.7$, eccentricity $e = 0.87$, periastron longitude $\omega = 138.7^\circ$, 
and inclination angle $i = 35^\circ$. The disk inclination angle is assumed to be $i_{\mathrm{d}}=10^{\circ}$. The disk is assumed to be 
perpendicular to the major axis of the orbit with a constant width of $10^{12}$\,cm (corresponds to $1^\circ$ half-opening angle at a 
distance of $45R_\ast$) and constant energy density of $8$\,erg\,cm$^{-3}$ (the highest energy density for which the \fermi\ upper limits are 
not violated, see previous subsection). The expected light curve after the total absorption assuming a constant initial \gammaray\ flux is shown 
in Fig. \ref{tev_lc} by the dashed line\footnote{Note, that there was a mistake in calculation of the angle between the disk normal and line of sight in the original paper \cite{2014JHEAp...3...18S}. Corrected results presented here show a slightly stronger effect of the disk absoprtion.}. The solid line represents the flux after absorption taking into account only stellar photons as reported 
by \cite{2006A&A...451....9D}. However, the assumption of the constant optical depth underestimates the effect of the \gaga-absorption in the disk as 
the optical depth $\tau_{\gamma\gamma,\,\mathrm{d}}(\epsilon)$ peaks at $\sim300$\,GeV. The resulting light curve after absorption in the disk only taking into 
account dependence of the optical depth on energy is shown by the dotted line in Fig. \ref{tev_lc}.  

It can be seen that the circumstellar disk might play a crucial role in the flux variability around the periastron passage and that the observed flux variability 
can be explained by the absorption. The time shift of the observed preperiastron peak with respect to the modelled light curve might be explained 
by the rotation of the disk normal with respect to the major axis.
The assumption of the constant width of the disk results in an overestimation 
of the absorption close to periastron, since the disk should be thinner close to the star. However, the energy density in the disk is expected to be highest close to the star. The combined effect might result in a shallower minimum near periastron than predicted by our calculations. 
The light curve has been modelled under the assumption that the primary \gammaray\ 
flux is constant. This is a valid approximation only in the case when IC scattering occurs in the saturation regime which is believed to be true around 
periastron. Farther from periastron the primary flux is expected to decrease. In the case when IC scattering is not in the saturation regime, local 
maxima are expected at the disk crossings and at the periastron. This would completely change the expected light curve. With the current quality 
of the observed TeV light curve it is difficult to draw any final conclusions. New observations conducted in 2014 might bring some 
clarity and give answers to some of these open questions. 

A critical test of the absorption model could be derived from the study of the spectral variability of the TeV emission. The TeV spectrum 
is expected to soften when the pulsar is moving through the disk for the first time, then the spectral shape should remain 
almost unchanged with a slight variability due to the absorption in the stellar radiation field until 
the pulsar moves out of the disk after periastron. After that the spectrum should harden again. Unfortunately, the VHE $\gamma$-ray 
observations during previous periastron passages did not provide enough statistics to draw firm conclusions about the spectral variability. 
Although there are some indications of 
spectral changes both during the 2004 \cite{2005A&A...442....1A} and the 2007 \cite{2009A&A...507..389A} periastron passages,  these 
changes are not significant. However, if the spectral variability is real on both occasions, it suggests a hardening of the spectrum 
towards periastron, which is in contradiction with the effect we expect in the case of effective cascading.

The study of the \gaga-absorption in the disk which would take into account a proper geometry of the disk and a realistic distribution of IR photons 
within the disk corresponding to the observed IR emission from the star is currently in progress.

%% \bibliographystyle{JHEP} 
%% \bibliography{references}

%% \begin{thebibliography}{99}
%% \bibitem{...} 
%% ....
\providecommand{\href}[2]{#2}\begingroup\raggedright\endgroup

%% \end{thebibliography}

\end{document}